\documentclass[twocolumn, trackchanges]{aastex701}
\usepackage{multirow}
\usepackage{amsmath}

\begin{document}

\title{Anisotropic Diffusion in Pulsar Halos: Interpreting the asymmetric morphology of Geminga and Monogem halos measured by HAWC}

\author[orcid=0009-0003-9828-6069,gname='Sizhe',sname='Wu']{Si-Zhe Wu}
\affiliation{School of Astronomy and Space Science, Nanjing University, Nanjing 210023, China} 
\affiliation{Key laboratory of Modern Astronomy and Astrophysics, Nanjing University, Ministry of Education, Nanjing 210023, China} 
\email{221840235@smail.nju.edu.cn}

\author[orcid=0000-0002-9654-9123, gname='Chaoming', sname='Li']{Chao-Ming Li}
\affiliation{School of Astronomy and Space Science, Nanjing University, Nanjing 210023, China} 
\affiliation{Key laboratory of Modern Astronomy and Astrophysics, Nanjing University, Ministry of Education, Nanjing 210023, China} 
\affiliation{Deutsches Elektronen Synchrotron (DESY), Platanenallee 6, D-15738 Zeuthen, Germany}
\email{chao-ming.li@desy.de}

\author[orcid=0000-0003-1576-0961,gname='Ruoyu',sname='Liu']{Ruo-Yu Liu}
\affiliation{School of Astronomy and Space Science, Nanjing University, Nanjing 210023, China} 
\affiliation{Key laboratory of Modern Astronomy and Astrophysics, Nanjing University, Ministry of Education, Nanjing 210023, China} 
\email{ryliu@nju.edu.cn}

\correspondingauthor{Ruo-Yu Liu}
\email{ryliu@nju.edu.cn}

\begin{abstract}
Pulsar halos are produced by electrons and positrons diffusing in the interstellar medium around their parent pulsar wind nebulae. Recent observations by HAWC and LHAASO have revealed asymmetric morphologies in the halos surrounding Geminga and Monogem. The anisotropic diffusion model provides a natural explanation for such asymmetries, where the morphology is determined by the viewing angle of the mean magnetic field, the Alfvénic Mach number ($M_{\rm A}$), and the pulsar distance. In this work, we model the measured morphologies based on this framework and constrain the properties of interstellar magnetic turbulence. We find that the mean magnetic field orientations within the two halos are different, implying that they reside in different magnetic coherence regions, whereas the Alfvénic Mach numbers are relatively close ($M_{\rm A}\sim 0.2$). The results suggest a local magnetic field coherence length of approximately 100\,pc. Our study demonstrates that the morphology of pulsar halos serves as a powerful diagnostic tool for the properties of interstellar magnetic fields, highlighting the need for more accurate morphological measurements and sophisticated diffusion modeling in future studies.
\end{abstract}


\section{Introduction} 
The detection of extended TeV emission surrounding the pulsars Geminga and Monogem by the High Altitude Water Cherenkov (HAWC) Observatory in 2017 revealed a new class of gamma-ray sources known as pulsar halos or TeV halos \citep{science_HAWC}. Subsequently, a growing number of similar sources (or candidates) have been detected by various instruments \citep{LHAASO_J0621, HAWC2023_J0359, LHAASO2025_J0248, HESS2023_Geminga}. These halos are understood to originate from high-energy electrons and positrons (hereafter electrons) that have escaped their parent pulsar wind nebulae (PWNe) and are diffusing through the ambient interstellar medium (ISM). These electrons produce extended gamma-ray emission via inverse Compton (IC) scattering off interstellar radiation fields, such as the cosmic microwave background (CMB) and infrared photons \citep{Liu2022, Lopez-Coto2022, Fang2022, Amato2024}.

A notable feature of these halos is the inferred slow diffusion of particles within them. By fitting the azimuthally-averaged one-dimensional (1D) radial surface brightness profiles (SBPs) of Geminga and Monogem, HAWC derived a diffusion coefficient of $D\sim 10^{28}\mathrm{cm^2/s}$ at 100\,TeV. This value is approximately two orders of magnitude lower than the typical Galactic diffusion coefficient derived from cosmic ray measurements ($D\sim 10^{30}\mathrm{cm^2/s}$ at 100\,TeV) \citep{Trotta_2011}. This discrepancy suggests that the transport of TeV electrons in the surrounding medium of these pulsars is highly suppressed compared to the average ISM.

To explain the low diffusion coefficient in pulsar halos, several models have been put forward. One scenario posits a region of highly amplified turbulence extending tens of parsecs around the pulsar, within which the diffusion coefficient is uniformly low due to the highly turbulent medium \citep{PositronExcess_Hooper_2017, Profumo2018, Fangkun19, Johan2019, Tang2019}. \citet{quasi-ballistic_propagation} suggested that the transport of electrons within pulsar halos is quasi-ballistic instead of diffusive at small distance, which avoids the necessity for an extremely low diffusion coefficient but instead demands a huge amount of total energy in injected electrons (see also critiques by \citealt{Bao2022}). A third model, the anisotropic diffusion model, offers a different perspective and does not invoke a highly turbulent environment either \citep{Liu_Anisotropic_Model}. It proposes that particle transport is governed by sub-Alfv\'enic turbulence in the ISM with a well-defined mean field direction. In this framework, particles diffuse preferentially along the mean magnetic field, while cross-field propagation is strongly suppressed. The slow effective diffusion inferred from azimuthally-averaged profiles can then be explained if the mean magnetic field is aligned close to the line of sight (LOS) of observers, thus projecting the slow, perpendicular diffusion onto the plane of the sky. 
While it is challenging to distinguish the three models solely by the radial SBP with currently measured accuracy \citep{Wu2024}, the anisotropic diffusion model predicts an asymmetric 2D morphology of the halo if the mean magnetic field direction is not closely aligned with the LOS \citep{Liu_Anisotropic_Model, Yan_LHAASO_prospect, Luque2022}. In contrast, the other models generally predict a more symmetric morphology, unless additional assumptions about a non-radial gradient in the diffusion coefficient are invoked (\citealt{Fang2025}).

Recent observations have provided interesting insights for morphological asymmetry. The Large High Altitude Air Shower Observatory (LHAASO) \citep{Chen2023Geminga}, and, more quantitatively, HAWC \citep{2024HAWC_Observation} have reported asymmetric morphology for both Geminga and Monogem halos. The HAWC study divided the region around each pulsar into four quadrants based on right ascension and declination, and independently fitted an isotropic diffusion model to the 1D SBP of each quadrant. The resulting best-fit diffusion coefficients showed a $\sim2\sigma$ level variation among the quadrants, providing suggestive support for an anisotropic diffusion scenario. This observed asymmetry may be used as a test for the anisotropic diffusion model. Also, as demonstrated by \citet{Li2026}, the asymmetric morphology of a pulsar halo can serve as a powerful probe of the local interstellar magnetic turbulence under the anisotropic diffusion framework, allowing us to diagnose key properties such as the Alfv\'enic Mach number ($M_{\rm A}$) and the orientation of the mean magnetic field. These findings motivate a more detailed modeling effort.

In this paper, we model the asymmetric morphologies of the Geminga and Monogem halos as measured by HAWC within the framework of the anisotropic diffusion model. By fitting the model predictions to the quadrant-dependent diffusion coefficients reported by \citet{2024HAWC_Observation}, we aim to constrain the properties of the local interstellar magnetic field in the vicinity of these two nearby pulsars. The remainder of the paper is structured as follows: In Section \ref{Methods}, we will introduce the methodology employed in this work, including the models for electron injection, energy loss, and the coordinate system we use. A detailed description of the anisotropic diffusion model will also be provided. In Section \ref{Result} and Section \ref{Discussion}, we present and discuss the main results of this work. And we will give our conclusion in Section \ref{Conclusion}.

\section{Methods}\label{Methods}
In the anisotropic diffusion model, the transport equation of electrons within the halo can be given by:
\begin{equation}\label{eq:diff}
\begin{split}
\frac{\partial N_e}{\partial t} = &
\frac{D_\perp  (E_e)}{r} \frac{\partial}{\partial r}
\left(r \frac{\partial N_e}{\partial r}\right)
+ D_\parallel (E_e) \frac{\partial^2 N_e}{\partial z^2}\\
& - \frac{\partial}{\partial E_e}(N_e \dot{E}_{\rm e})
+ Q(E_e, t) \delta(\vec{r})
\end{split}
\end{equation}
where the $z$-axis is defined as the mean magnetic field direction. Here $N_e$ is the differential number density of electrons, $D_\parallel$ and $D_\perp$ are the diffusion coefficients of particles along and perpendicular to the mean magnetic field direction, respectively. They are related by $D_\perp=M_{\rm A}^4M_{\parallel}$ \citep{Yan_2008}. The parallel diffusion coefficient is assumed to be $D_\parallel(E_e)= D_0\left({E_e}/{1\text{GeV}} \right)^\delta$, with $\delta$ being set to $1/3$ and $D_0$ being set to $4\times10^{28}\mathrm{cm^2/s}$ following the standard value for ISM \citep{Trotta_2011}. $\dot{E}_{\rm e}$ represents the energy loss rate of injected electrons, which is primarily contributed by synchrotron radiation and IC radiation. It is given by
\begin{equation}
    \dot{E}_{\rm e} =-\frac{4}{3}\sigma_Tc\left(\frac{E_e}{m_ec^2}\right)^2\left [  U_B+\sum_{i}f_{KN}(E_e,T_i)U_i \right ]
\end{equation}
where $\sigma_T$ is the Thomson cross section, $U_B$ and $U_i$ are the energy densities of magnetic field and photon field, respectively. The function $f_{KN}(E_e,T)$ represents the influence of the Klein-Nishina effect\citep{cao2021_fKN}:
\begin{equation}
    f_{KN}(E_e, T_i)= \left [ 1+\left(\frac{2.82k_BT_iE_e}{m_e^2c^4}\right)^{0.6}\right ]^{-\frac{1.9}{0.6}}
\end{equation}
We set the magnetic field strength to $B=3\, \mu$G (corresponding to a magnetic field energy density of 0.22\,eV/cm$^3$), and consider the CMB ($U_{\rm CMB}=0.26\text{ eV/cm}^3,\ T_{\rm CMB}=2.73
\text{ K}$) and far-infrared ($U_{\rm FIR}=0.25\text{ eV/cm}^3,\ T_{\rm FIR}=30
\text{ K}$) blackbody/graybody radiation as the background photon fields, which are similar to that employed in \citet{2024HAWC_Observation}. The hotter photon field can be neglected due to the suppression by the Klein-Nishina effect.

The injection term in Eq.~(\ref{eq:diff}), $Q(E_e, t)$,  is assumed to follow the spindown evolution of the pulsar, i.e.,
\begin{equation}
    Q(E_e, t)=Q_0 \left(1+\frac{t}{\tau_0}\right)^{-2}E_e^{-\alpha}\exp\left(-\frac{E_{\rm e}}{E_{\rm cut}}\right)
\end{equation}
where the typical pulsar spin-down timescale $\tau_0$ is assumed to be 12\,kyr, the spectral index $\alpha$ is set to 1 and cutoff energy $E_{\rm cut}$ is set to 100 TeV, which basically follow the results of \citet{2024HAWC_Observation}. The normalization factor $Q_0$ is obtained by $\int E_{\rm e}Q(E_{\rm e},t_{\rm age})dE_{\rm e}=\eta_{\rm e}L_{\rm s}(t_{\rm age})$ with $t_{\rm age}$ being the age of the pulsar and $L_{\rm s}(t_{\rm age})$ being the current spindown luminosity of the pulsar as inferred from pulsar timing measurement. $\eta_{\rm e}$ is a factor parameterizing the fraction of the spindown energy of the pulsar going to the energy of relativistic electrons escaped to the surrounding ISM. Since the size of a PWN is significantly smaller than that of the pulsar halo, we treat the injection of electrons as a point injection, reflected by the $\delta$ function in the last term.

\begin{figure}[t]
    \centering
    \includegraphics[width=0.9\linewidth]{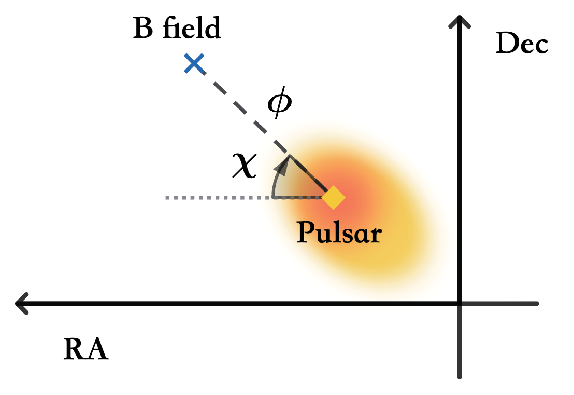}
    \caption{Sketch figure for the definition of $\phi$ and $\chi$. The location of $B$-field represents the projection of the mean magnetic field direction on the celestial sphere. Please refer to the appendix for the formula used to calculate the equatorial coordinates of the magnetic field direction.}
    \label{fig:coorsys}
\end{figure}

\begin{table*}[t]
    \centering
    \setlength{\tabcolsep}{18pt}
    \begin{tabular}{cccccc}
    \hline
    \multirow{2}{*}{Parameter} & \multicolumn{2}{c}{Geminga} & & \multicolumn{2}{c}{Monogem}\\
    \cline{2-3} \cline{5-6}
    & {$r_{\rm max}=100$ pc} & $r_{\rm max}=200$ pc & & $r_{\rm max}=100$ pc & $r_{\rm max}=200$ pc \\
    \hline
    $\phi\ [^\circ]$ & $19^{+25}_{-11}$ & $3^{+19}_{-1}$ & & $29^{+8}_{-13}$ & $11^{+6}_{-7}$\\
    $\chi\ [^\circ]$ & $-6^{+30}_{-11}$ & $15^{+20}_{-32}$ & & $82^{+4}_{-41}$ & $59^{+22}_{-16}$\\
    $M_{\rm A}$ & $0.23^{+0.09}_{-0.05}$ & $0.23^{+0.12}_{-0.02}$ & & $0.16^{+0.09}_{-0.02}$ & $0.22^{+0.03}_{-0.03}$\\
    $d$ [pc] & $280^{+157}_{-69}$ & $267^{+275}_{-17}$ & & $282^{+57}_{-39}$ & $317^{+50}_{-38}$\\
    $\eta_e$ & $0.20^{+0.27}_{-0.10}$ & $0.07^{+0.44}_{-0.02}$ & & $0.12^{+0.05}_{-0.03}$ & $0.07^{+0.03}_{-0.02}$ \\
    \hline
    \end{tabular}
    \caption{Best-fit values and the 90\% confidence intervals obtained under the anisotropic diffusion model for Geminga and Monogem.}
    \label{tab:bestfit}
\end{table*}

\begin{figure*}[t]
    \centering
    \includegraphics[width=0.95\linewidth]{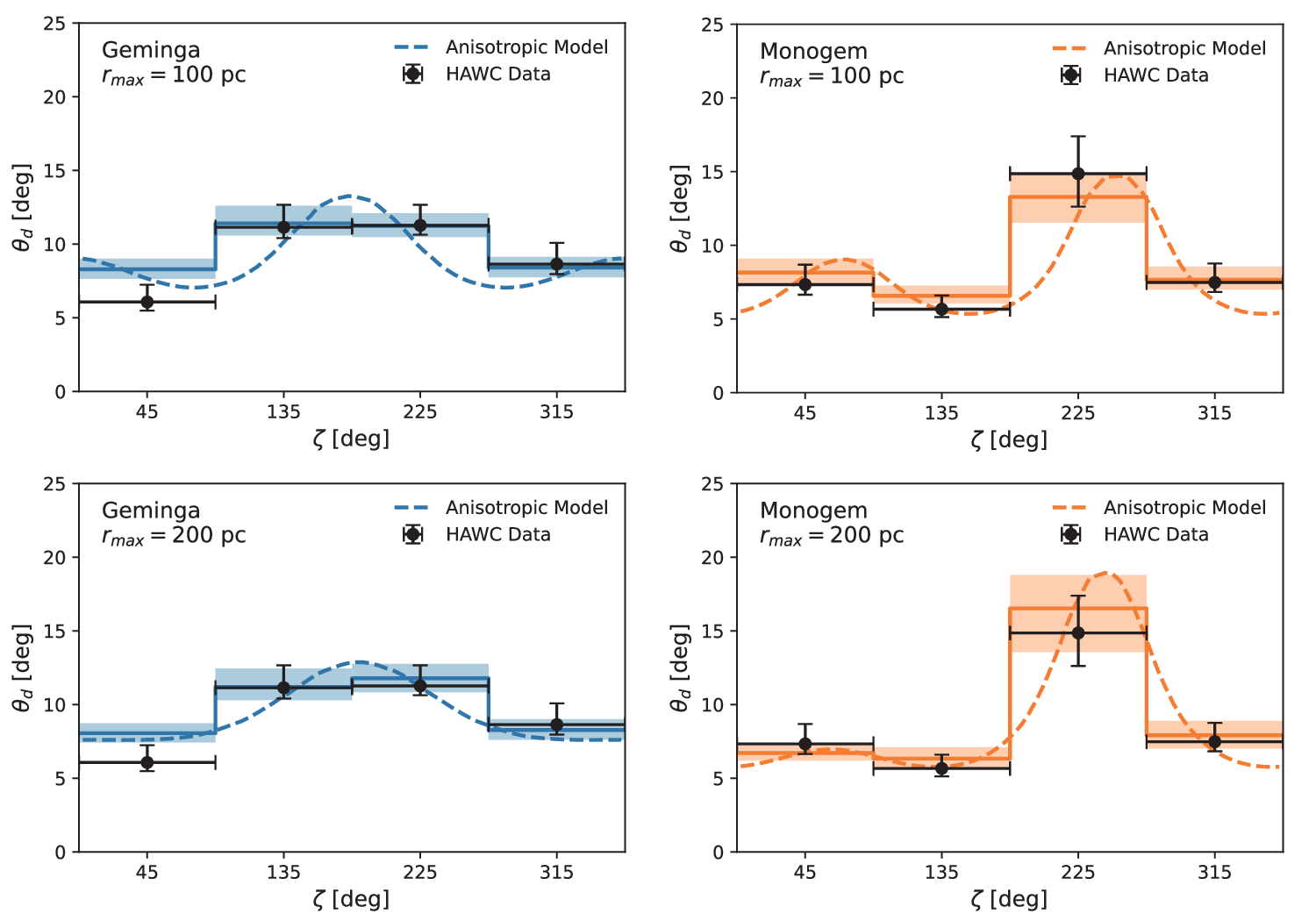}
    \caption{Variation of the characteristic diffusion angle $\theta_d$ with azimuth $\zeta$ for Geminga and Monogem with different integration radius. The colored dashed line represents the characteristic diffusion angle at each azimuth, calculated from the median of the posterior distribution of the anisotropic model parameters. The solid colored line and the shaded region indicate the median and the 68\% error range fitted for each quadrant, respectively.}
    \label{fig:Azimuth_distribution}
\end{figure*}

Solving this diffusion equation, we obtain the 3D spatial distribution of electron density around the two pulsars at the current time (i.e., $t_{\rm age}=370$\,kyr for Geminga and 110\,kyr for Monogem). 

Then we can calculate the $\gamma$-ray emissivity and integrate it along the LOS to obtain a 2D intensity map, based on which we can obtain the radial SBP along any direction. The SBP of the halo measured by HAWC extends to about 10 degrees away from the pulsar, corresponding to approximately 50\,pc given the nominal distance of 250\,pc. We therefore set the integration region as a sphere with a radius $r_{\rm max}$ centered on the pulsar, considering $r_{\rm max}=100\,$pc and 200\,pc separately in the later calculation. Following the observation energy range of HAWC \citep{2024HAWC_Observation}, we integrate $\gamma$-ray photons from 1 to 316\,TeV in this work. 

Fig.~\ref{fig:coorsys} shows the coordinate system we use in this work. The location of $B$-field (the blue cross) represents the direction on the celestial sphere towards which the local mean magnetic field within the halo points. $\phi$ represents the angular distance between the pulsar and the projection of $B$-field in celestial sphere (i.e., the inclination angle between the mean field direction and the LOS), and azimuth $\chi$ represents the direction of the offset between the mean magnetic field and the LOS relative to the positive RA direction. 

\citet{2024HAWC_Observation} did not provide the average SBP within each quadrant, but instead reported the best-fit effective diffusion coefficient of electrons normalized at 100\,TeV within each quadrant under the isotropic diffusion model. It has been suggested that the SBP predicted by the isotropic diffusion model can be described with the formula $I(\theta)\propto \theta_d^{-1}(\theta+0.06\theta_d)^{-1}\exp(-\theta^2/\theta_d^2)$ with $\theta$ the angle from the source to a certain point in the sky \citep{science_HAWC}. $\theta_d(E)$ is a characteristic parameter which is related to the diffusion coefficient by $\theta_d(E)=2\sqrt{D(E)t_{\rm e}}/d$, with $t_{\rm e}$ being the smaller one between the age of the pulsar and the cooling timescale of electrons primarily producing gamma rays of energy $E$. Since the magnetic field and the radiation fields are already given, the reported diffusion coefficient $D(100\,\rm TeV)$, or equivalently, $\theta_d$, can describe the average SBP within each quadrant. At this characteristic angular distance, the intensity drops to  $I(\theta_d)/I(0)\approx 0.021$ of the central value. Therefore, to compare the 2D intensity map $I_{2D}(\theta, \zeta)$ predicted by the anisotropic diffusion model to the observation of HAWC, we also divide the predicted 2D map into four quadrants, obtain the average SBP within each quadrant, and look for the angular distance where the intensity drops to 0.021 times of the central intensity. As such, we obtain $\theta_{d}$ (or the effective diffusion coefficient) from model, which satisfies 
\begin{equation}
\frac{\int_{\zeta_1}^{\zeta_2} I(\theta_d,\zeta)\mathrm{d}\zeta}{\int_{\zeta_1}^{\zeta_2}I(0,\zeta)\mathrm{d}\zeta}=0.021
\end{equation}
with $\zeta_1$ and $\zeta_2$ being the azimuthal angles at which a quadrant starts and ends respectively\footnote{\cite{2024HAWC_Observation} divides the halo into four quadrants based on the right ascension and declination of the pulsar's position. However, as declination follows small circles, this method results in uneven shape between the northern and southern quadrants. For simplicity of our calculation, we divide the quadrants by the great circle tangent to the parallel of latitude at the pulsar's location and the meridian passing through that position so that the shape of each quadrant is identical (i.e., a right-angle spherical sector). Given the low declinations of the two pulsar halos ($\lesssim 20^\circ$), the deformation in the quadrant shapes under the HAWC definition is minimal, and the difference between the two quadrant definitions is also negligible.}. We then can adjust parameters for the local magnetic field such as $\phi$, $\chi$ and $M_{\rm A}$ to match the predicted $\theta_d$ with the measured $\theta_d$. 

\section{Result}\label{Result}

\begin{figure*}[t]
    \centering
    \includegraphics[width=0.95\linewidth]{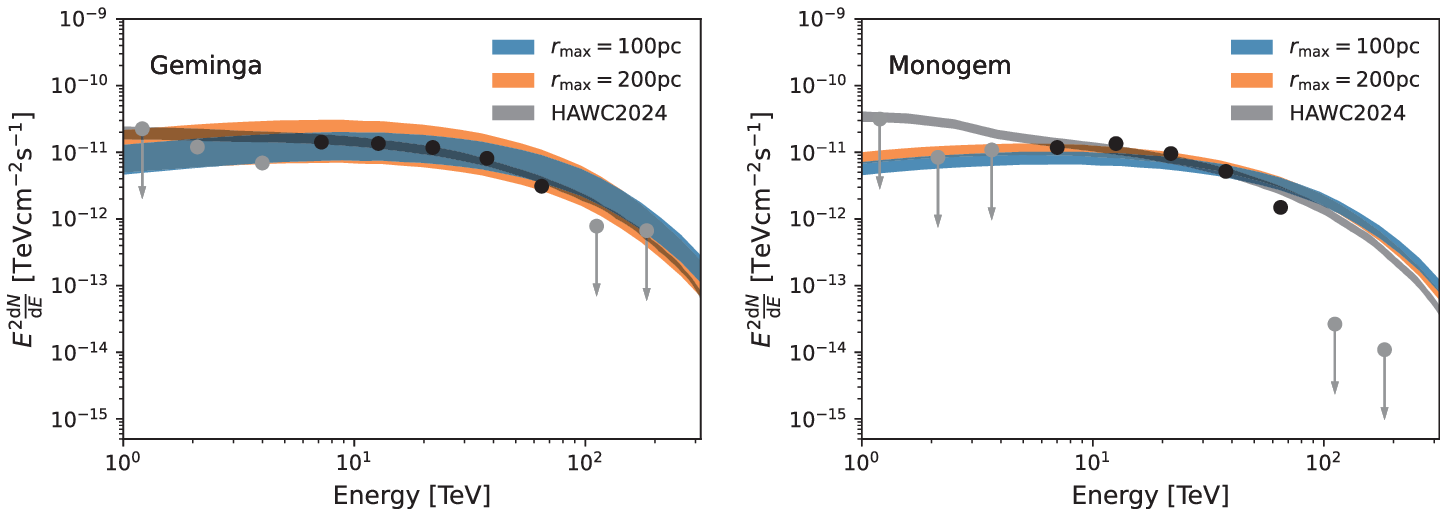}
    \caption{Spectra fitting to the spectra of Geminga (left panel) and Monogem (right panel), with blue and orange bands being uncertainty bands of 68\% confidence intervals for the cases of $r_{\rm max}=100\,$pc and 200\,pc respectively. Gray bands show for comparison the results obtained by HAWC with the diffusion template \citep{2024HAWC_Observation}. The black data points indicate the data used in the $\eta_e$ fit, while the gray data points are shown for reference only and not included in the fitting.}
    \label{fig:Spectrum}
\end{figure*}

Following the above procedure, we perform a scan over the parameter space with the Monte-Carlo Markov Chain (MCMC) method, employing the Python package {\tt emcee}\citep{Foreman_2013}. In addition to 
$\phi$, $\chi$, and $M_{\rm A}$, the distances of Geminga and Monogem from Earth ($d_G$ and $d_M$) are also treated as free parameters given the significant uncertainties in these distance measurements ($d_G=250^{+120}_{-62}\text{ pc}$, and $d_M=288^{+33}_{-27}\text{ pc}$) \citep{Pulsar_distance_Manchester}. 

As discussed in \citet{Li2026}, if the distance between the pulsar and the observer is significantly greater than the size of the pulsar halo, there exists a degeneracy between $\phi$ and $M_{\rm A}$, and the pulsar halo would appear as an ellipse with axial symmetry. In contrast, for Geminga and Monogem, due to their proximity to Earth, the halo size is comparable to the distance, so the predicted halo morphology appears cometary shape for $\phi> 0$ \citep{Liu_Anisotropic_Model, Fang_Wing_Morphology}, producing different extensions among all four quadrants. This breaks the degeneracy between $\phi$ and $M_{\rm A}$.

Fig.~\ref{fig:Azimuth_distribution} shows the fitting to $\theta_d$ with the anisotropic diffusion model, in the cases of $r_{\rm max}=100\,$pc and $r_{\rm max}=200\,$pc respectively, and the best-fit parameters are shown in Table~\ref{tab:bestfit}\footnote{We have excluded one of the solutions obtained from the MCMC fitting for Monogem with $r_{\rm max}=200$ pc. Please refer to the appendix for more details.}. The best-fit values of the mean magnetic field orientation in two cases of $r_{\rm max}$ are different, although the differences are not very significant given the large uncertainty ranges. We note that, under our adopted cooling and diffusion parameters, 100\,TeV electrons can propagate several hundred parsecs along the mean magnetic field before cooling, which exceeds the integration radii (i.e., $r_{\rm max}$) adopted in the calculation. Electrons beyond $r_{\rm max}$ may in principle contribute to the total emission. We therefore test the influence of employed $r_{\rm max}$ by extending it to larger values, and list the corresponding fitting results in Appendix~\ref{sec:appendixC}. We find that the inferred magnetic-field orientation and $M_{\rm A}$ vary only moderately with increasing $r_{\rm max}$, and our main conclusions remain unchanged. This is because the electron density decreases with distance from the pulsar, and a considerable portion of
the emission of electrons beyond $r_{\rm max}$ is located outside the region of interest of HAWC's analysis (which is about $10^\circ$), in particular when $\phi$ is remarkably greater than $0^\circ$. It should be also pointed out that the integration radius $r_{\rm max}$ introduced in our calculations is not intended to contain contribution of all injected electrons. Rather, it is introduced to describe the dominant gamma-ray emission projected within the HAWC region of interest under the assumption that the halo is embedded in a single magnetic-coherence domain. Given the typical coherence length of the interstellar magnetic field is not expected to be much larger than a few hundreds of parsecs, employing a larger $r_{\rm max}$, which is equivalent to assuming a coherence length $\geq 2r_{\rm max}$, would make the single-coherence approximation less self-consistent.

%, which would not affect the fitting of $\theta_d$. We calculated the ratio of flux within 200\,pc to the total flux, considering the best-fit value of model parameters, and a field of view radius of 10 degrees (i.e., approximately the same with the field of view adopted by HAWC and the value of $\theta_d$). The fractions are approximately 85\% and 95\% for Geminga and Monogem, respectively. This indicate that an integration radius of 200\,pc is sufficient to contain most of the flux. Adopting a larger integration radius implies that we assume a larger-scale magnetic coherence, we will discuss it further in Section~\ref{Discussion}, and the fitting results with larger integration regions are also provided in the Appendix, showing not very significant difference with that of $r_{\rm max}=200\,$pc.} %We have checked the results with larger $r_{\rm max}$, and the results are similar to that of $r_{\rm max}=200\,$pc. The influence of $r_{\rm max}$ will be discussed further in Section~\ref{Discussion}.

In the previous study, the inclination angle $\phi$ for Geminga is suggested to be less than $5^\circ$ \citep{Liu_Anisotropic_Model, Luque2022}. This is because the asymmetric morphology of the halo was not revealed at the time and $\phi<5^\circ$ is employed to keep quasi-isotropy of the halo's morphology. It led to a tension between the non-detection of elongated morphology of halos and the small probability (i.e., approximately 0.4\%) of viewing angle being within $5^\circ$. In this study, the fitted value of $\phi$ is much larger than the previous estimate, significantly relaxing the tension. On the other hand, $M_{\rm A}$ obtained in the two cases are generally consistent, which are around 0.2 for both Geminga and Monogem, and are compatible with previous studies, as well as with an independent method with radio polarization analysis \citep{Pavaskar2023}.

For electrons with energies above 100\,TeV diffusing along the mean magnetic field, the scattering mean free path is approximately 
$\sim 50$\,pc under the employed diffusion coefficient. This suggests that the directions of recently injected high-energy electrons may not yet be fully randomized, such that their transport is not accurately described by the diffusion approximation \citep{Prosekin2015,quasi-ballistic_propagation, pu2026}. To assess the impact of this effect, we exclude electrons for which the elapsed time since injection is shorter than the isotropization timescale $\tau\approx3D_\parallel/c^2$, and refit the model. We find that the resulting change in the best-fit model parameters is negligible.

Based on the $\gamma$-ray spectrum measured by HAWC in the range of approximately $5$–$85$\,TeV, we estimated the electron injection energy efficiency $\eta_e$ for Geminga and Monogem in Table~\ref{tab:bestfit} by fitting the spectrum in Fig.~\ref{fig:Spectrum}. The uncertainties of $\eta_e$ mainly arise from the uncertainties of distances of pulsars. Additionally, due to the adoption of the finite integration radius $r_{\rm max}$, electrons with relatively lower energy may diffuse beyond the integration region, thereby affecting the estimate of $\eta_e$ and the $\gamma$-ray spectrum.

We also examined the influence of $D_0$ by increasing $D_0$ from $2\times10^{27}\mathrm{cm^2/s}$ to $6\times10^{28}\mathrm{cm^2/s}$. The viewing angle $\phi$ and azimuthal angle $\chi$ of magnetic field remain essentially unchanged under different diffusion coefficients. The best-fit value of Alfv\'enic Mach number $M_{\rm A}$ decreases with increasing diffusion coefficient, approximately following the relation $M_{\rm A}\propto {D_0}^{-\frac{1}{4}}$. Combined with $D_\perp=M_{\rm A}^4M_{\parallel}$, it can be concluded that the diffusion coefficient in the vertical direction $D_{\perp}$ remains essentially unchanged.

\section{Discussion}\label{Discussion}

\subsection{The distribution of magnetic field and the influence of coherence length}

In the preceding section, we modeled the asymmetric morphologies of the Geminga and Monogem pulsar halos separately, integrating the emission within $r_{\rm max}=100\,$pc and 200\,pc around each pulsar. Notably, the two pulsars are spatially proximate, separated by $\sim 100$\,pc. Consequently, their emission regions significantly overlap. Assuming a separation of 100 pc, we estimate that 31\% and 64\% of the integration volumes of the two halos overlap for $r_{\rm max}=100\,$pc and 200\,pc, respectively. Therefore, the inferred mean magnetic field orientation and $M_{\rm A}$ within the two halos are not supposed to show large differences. 

\begin{figure}[t]
    \centering
    \includegraphics[width=0.95\linewidth]{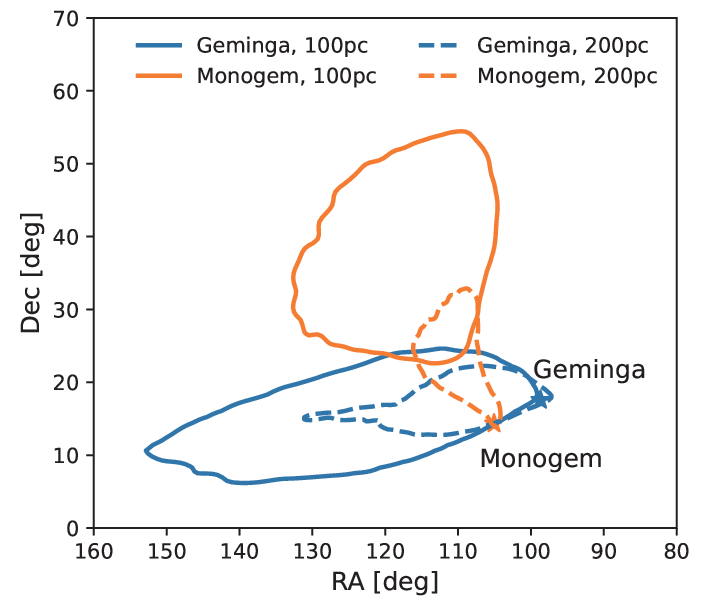}
    \caption{The 90\% confidence interval contour of the magnetic field direction distribution in the celestial sphere for Geminga and Monogem. The solid and dashed lines represent the results obtained with integration radii $r_{\rm max}$ of 100 pc and 200 pc, respectively.}
    \label{fig:mag_distribution}
\end{figure}

\begin{figure*}[t]
    \centering
    \includegraphics[width=0.95\linewidth]{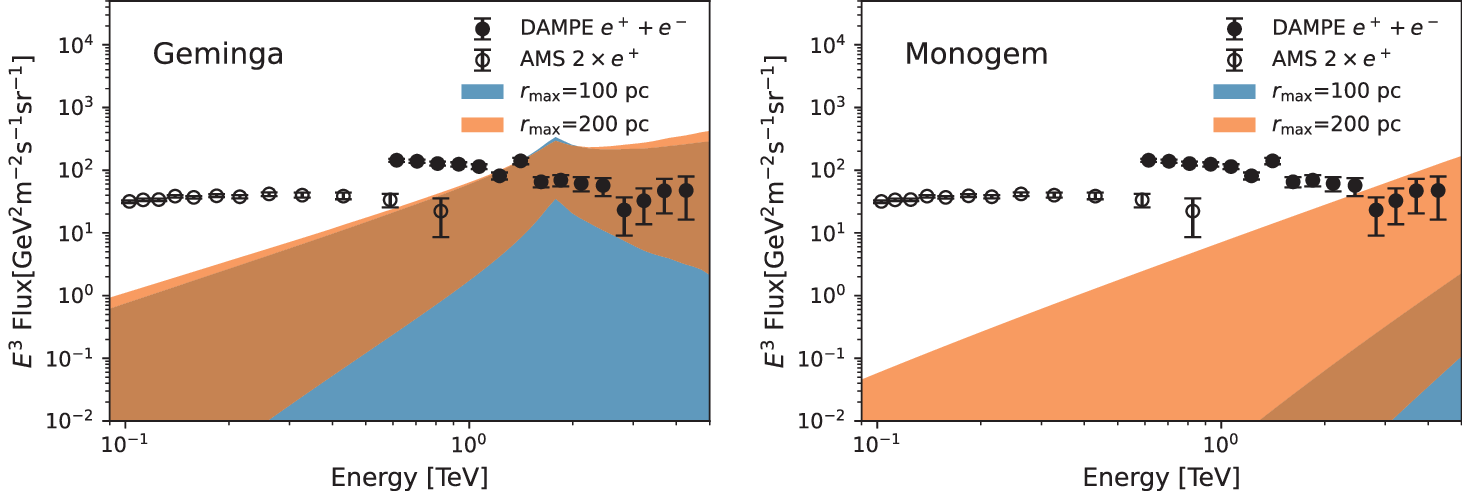}
    \caption{The 90\% confidence intervals of the contribution of Geminga and Monogem to electron and positron spectrum at Earth. The black solid and hollow data points data points represent the spectrum measured by DAMPE \citep{DAMPE_Observation} and AMS \citep{AMSPositron2019}, respectively.}
    \label{fig:local_e_spec}
\end{figure*}

Fig.~\ref{fig:mag_distribution} presents the 90\% confidence intervals for the mean magnetic field directions of Geminga and Monogem, derived from our MCMC fitting. For the case of $r_{\rm max}=100\,$pc, the angle between the best-fit magnetic field directions of Geminga and Monogem is about $29^\circ$, which is not very large. The confidence contours show minimal overlap, suggesting that the two halos may reside within different coherence of the magnetic field. For $r_{\rm max}=200\,$pc, the angle between the best-fit magnetic field directions of Geminga and Monogem decreases to about $10^\circ$, and the contours exhibit partial overlap. This is consistent with the larger integration volume, which encompasses a greater overlapping region and thus yields more similar inferred mean magnetic field properties. These results suggest that the coherence length of the local magnetic field is $L_{\rm c}\sim 100\,$pc. A shorter coherence length (e.g., $\sim 10\,$pc) would not induce obvious anisotropy. This was studied by \citet{Lopez-Coto2018} by performing a test-particle simulation in synthetic magnetic fields, and they found that for coherence length $L_{\rm c}=40\,$pc the predicted halo morphology shows asymmetry but the asymmetry becomes not obvious or disappears for $L_{\rm c}\leq 10\,$pc. %On the other hand, a significantly longer coherence length (e.g., $>500$\,pc) would result in identical mean field directions for the two halos, even for $r_{\rm max}=100\,$pc. It should be noted that our calculations only account for electrons within $r_{\rm max}$, neglecting those beyond this radius. For $D_0\sim 10^{28}\,\rm cm^2/s$,
On the other hand, a significantly longer coherence length (e.g., $>500$\,pc) would result in identical mean field directions for the two halos, even for $r_{\rm max}=100\,$pc.
As discussed in Section~\ref{Result}, the parallel diffusion length of 100\,TeV electrons could reach several hundreds parsecs, potentially crossing multiple magnetic coherence. Although the electron density decreases with distance, mitigating the impact of distant particles, their inclusion could complicate the halo morphology. Potential effects include a reduction in halo asymmetry \citep{Yan_Multiple_Coherence} and emergence of features such as ``wings'' \citep{Fang_Wing_Morphology, Yan_Multiple_Coherence} or ``mirage'' sources \citep{Bao_Mirage_sources_PRL, Bao_Mirage_sources}.

Reproducing the observed morphology of a specific pulsar halo under a multi-coherence scenario is challenging, as the outcome is highly sensitive to the stochastic orientation of the magnetic coherence. Also, scanning a sufficiently large parameter space to identify the best-fit model is computationally expensive. Moreover, relying solely on effective isotropic diffusion coefficients measured across four quadrants limits our ability to constrain the properties of multiple magnetic coherence regions. Future, more precise measurements of pulsar halo morphologies, coupled with a more sophisticated modeling of electron distribution under the multi-coherence scenario, will be essential for testing the anisotropic diffusion model and constraining the detailed properties of interstellar magnetic fields.

\subsection{The contribution of pulsar halos to the local electron spectrum}

Nearby middle-aged pulsars such as Geminga have long been regarded as important contributors to the local electron and positron spectrum, especially at high energies ($\gtrsim 0.1$\,TeV)\citep{AMSElectron2019, AMSPositron2019, DAMPE_Observation}, because particles accelerated in their pulsar wind nebulae (PWNe) may eventually propagate to the vicinity of the Earth \citep{Positron_Excess_Hooper_2009, Positron_Excess_Delahaye,Linden2013, Positron_Excess_Yin}. Motivated by this possibility, many previous studies have investigated the contributions of Geminga and Monogem, particularly Geminga, to the local electron/positron spectrum under different diffusion scenarios and observational constraints \citep{science_HAWC, TwoZoneModel, PositronExcess_Hooper_2017}. In the anisotropic diffusion scenario, however, such contributions become especially sensitive to the magnetic-field configuration, because the diffusion coefficients parallel and perpendicular to the mean magnetic field can differ substantially. For example, \citet{Xia_2025} estimated the contribution of Geminga for the special case of $\phi=0^{\circ}$, which leads to a relatively high predicted flux.

While we do not aim to explain the spectrum of electrons or positrons by Geminga and Monogem in this work, the expected flux from Geminga and Monogem are not supposed to exceed the measured flux, given the existence of other possible sources of positrons and electrons. This may serve as a constraint on the model. Based on the best-fit magnetic field properties obtained in our fitting, we estimate the contributions of Geminga and Monogem to the local electron/positron spectrum, as shown in Fig.~\ref{fig:local_e_spec}. In this calculation, we assume that the orientation of the magnetic field between the Earth and each pulsar is the same as the best-fit orientation inferred within the corresponding halo. Under this assumption, our model predicts a lower electron/positron flux than that obtained by \citet{Xia_2025}, mainly because our best-fit value of $\phi$ is larger than $0^{\circ}$. The predicted flux is below, or at most comparable to, the measured flux within the uncertainties.  Therefore, our model does not contradict with the present observations. We note, however, that these estimates depend on the assumed large-scale magnetic-field geometry. If multiple magnetic coherence exist between the Earth and the pulsars, the corresponding contributions to the local electron spectrum would become much more uncertain and difficult to predict.

\section{Conclusion}\label{Conclusion}
In this work, we have successfully modeled the asymmetric morphologies of the Geminga and Monogem pulsar halos reported by HAWC, utilizing an anisotropic diffusion framework. Our analysis allows for quantitative constraints on the properties of the interstellar magnetic turbulence within these halos, specifically the orientation of the mean magnetic field and the Alfvénic Mach number ($M_{\rm A}$). We find that the turbulence within both halos is sub-Alfv\'enic, with fitted $M_{\rm A}$ values consistently around 0.2. This result supports the scenario where the observed slow diffusion is primarily attributed to suppressed cross-field propagation. Furthermore, the robustness of the derived perpendicular diffusion coefficient suggests that our constraints on the turbulent properties are reliable across a reasonable range of assumed parallel diffusion coefficients.

The main finding of this study is the revised constraint on the inclination angle between the mean magnetic field and the line of sight. The fitted inclination angles are found to be significantly larger than previous estimates, which were derived under the assumption of quasi-isotropic morphology. This revision effectively resolves the statistical tension arising from the small probability of the line of sight aligning closely with the magnetic field direction. Additionally, the distinct mean magnetic field orientations inferred for Geminga and Monogem suggest that the two halos reside within different magnetic coherence regions. By comparing the differences in field directions derived from varying integration radii, we estimate that the coherence length of the local interstellar magnetic field is approximately 100 pc.

Finally, we note that the current single-coherence model represents a simplification of the complex interstellar environment. In a multi-coherence scenario, the morphology would be influenced by the stochastic configuration of magnetic domains, which is challenging to model with current data. In the future, high-precision observations of pulsar halo morphologies will be essential to further test the anisotropic diffusion model and to resolve the detailed structure of interstellar magnetic turbulence.

\begin{acknowledgments}
We would like to thank Kai Yan for helpful discussion. This work is supported by National Natural Science Foundation of China under grants No.~12393852 and 12333006, and Basic Research Program of Jiangsu under grant No.~BK20250059.
\end{acknowledgments}

\appendix
\section{Corner plots for MCMC fittings}
\begin{figure}[htbp]
    \centering
    \includegraphics[width=0.9\linewidth]{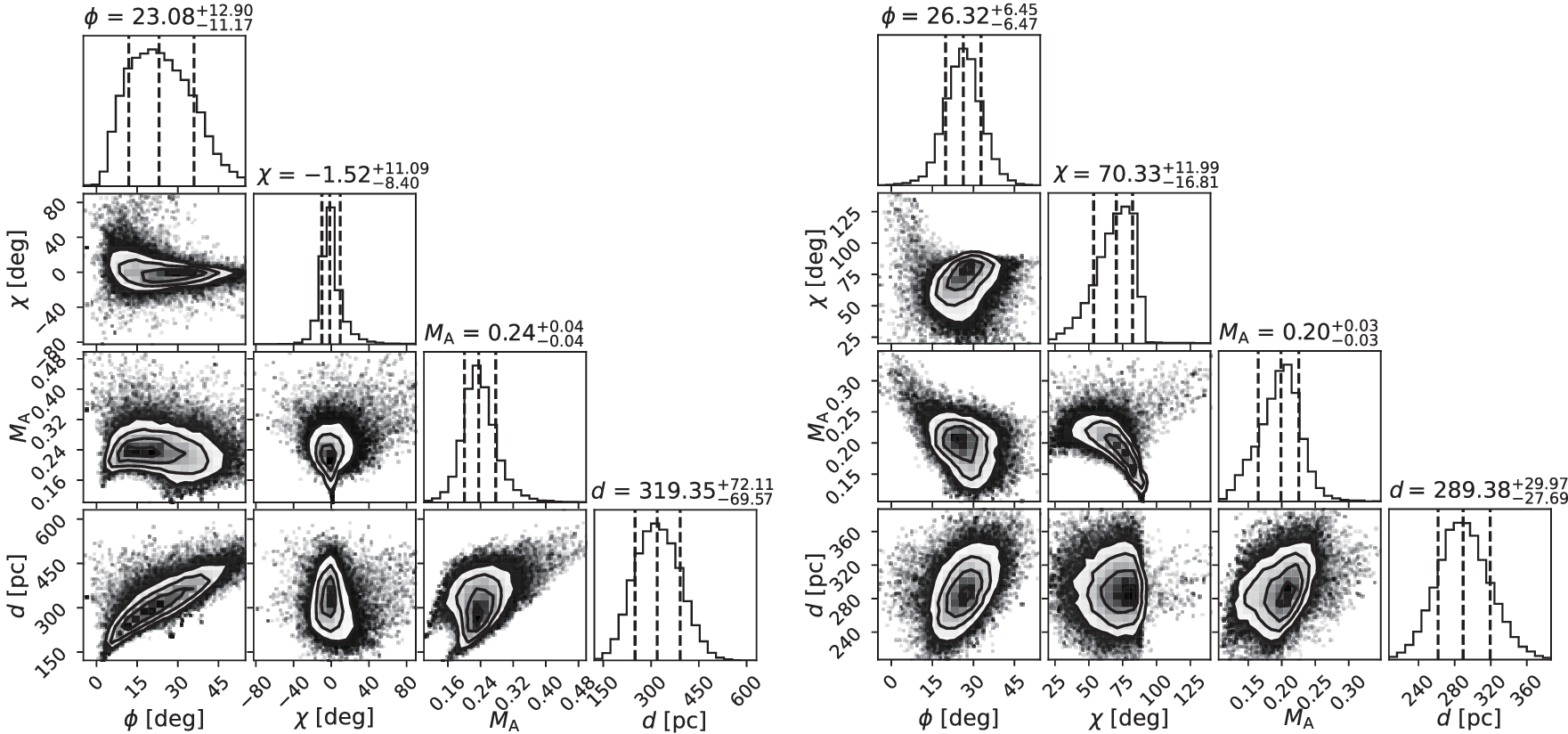}
    \caption{The corner plots for Geminga (left) and Monogem (right) with integration radius $r_{\rm max}=100$ pc obtained using the MCMC method.}
    \label{fig:MCMC_100}
\end{figure}

\begin{figure}[htbp]
    \centering
    \includegraphics[width=0.9\linewidth]{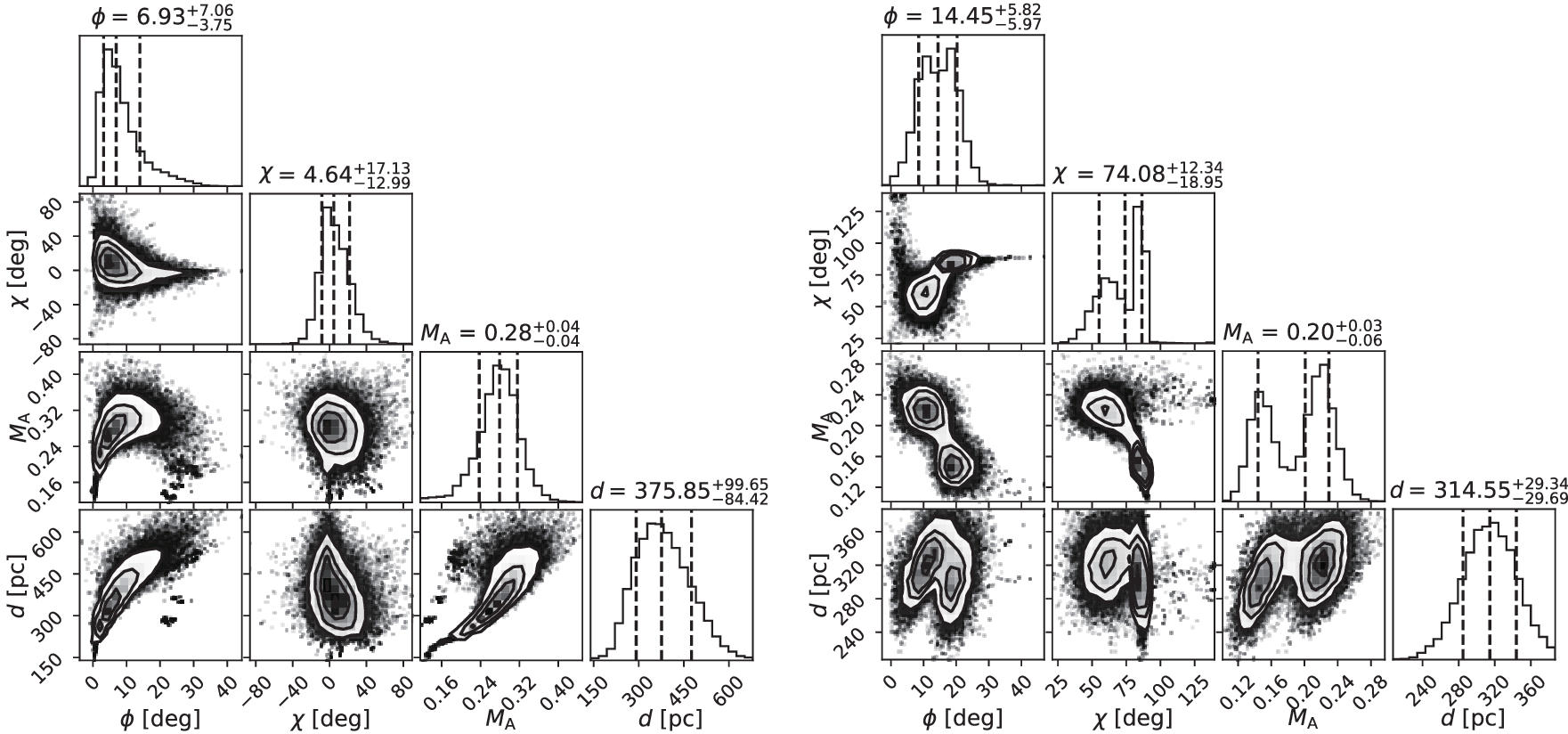}
    \caption{Same as Fig.~\ref{fig:MCMC_100}, but with $r_{\rm max}=200$ pc}
    \label{fig:MCMC_200}
\end{figure}

Corner plots for the MCMC fitting in the case of $r_{\rm max}=100\,$pc and $r_{\rm max}=200\,$pc are shown in Figures~\ref{fig:MCMC_100} and \ref{fig:MCMC_200}, respectively. The fitting results for Monogem with $r_{\rm max}=200\,$pc exhibit a bimodal distribution. For the peak with relatively smaller $M_{\rm A}$, $\theta_d$ in the third quadrant is obtained by averaging the extremely large ($\sim 35^\circ$) and extremely small values ($\sim 2^\circ$) of $\theta_d$ from different azimuthal angles. We did not find such significant asymmetry in the image observed by HAWC. In the main text, we exclude the peak with relatively smaller $M_{\rm A}$ value, and this does not affect our main conclusions.\\

\section{Calculation of Celestial Coordinates for Magnetic Field Direction}

We define the rotation matrix as Eq.~\ref{Eq:RotationMatrix}, where $\beta$ represents an arbitrary angle:
\begin{equation}
    P_x(\beta)=\begin{bmatrix}
 1 & 0 & 0\\
 0 & \cos\beta & -\sin\beta \\
 0 &  \sin\beta&\cos\beta
\end{bmatrix}
\,,\,
P_y(\beta)=\begin{bmatrix}
 \cos\beta & 0 & \sin\beta\\
 0 & 1 & 0\\
 -\sin\beta  & 0 &\cos\beta
\end{bmatrix}
\,,\,
P_z(\beta)=\begin{bmatrix}
 \cos\beta & -\sin\beta & 0\\
 \sin\beta & \cos\beta & 0\\
  0 & 0 & 1
\end{bmatrix}
\label{Eq:RotationMatrix}
\end{equation}

Then the vector of magnetic field direction can be written as:

\begin{equation}
    \vec{r_B}=P_y(\alpha_p)P_x(-\delta_p)P_z(\chi)P_y(\phi)
\begin{bmatrix}
0 \\
0 \\
1
\end{bmatrix}
\end{equation}

where $\delta_p$ and $\alpha_p$ are the declination and right ascension of the pulsar, respectively. Last, the declination ($\delta_B$) and right ascension ($\alpha_B$) of magnetic field direction can be written as:
\begin{equation}
    \delta_B=\arcsin(\vec{r_B}\cdot \vec{e_y})\,,\,\alpha_B=\arctan(\frac{\vec{r_B}\cdot \vec{e_x}}{\vec{r_B}\cdot \vec{e_z}})
\end{equation}

\section{The case of larger integration radius\label{sec:appendixC}}

\begin{figure}[htbp]
    \centering
    \includegraphics[width=0.5\linewidth]{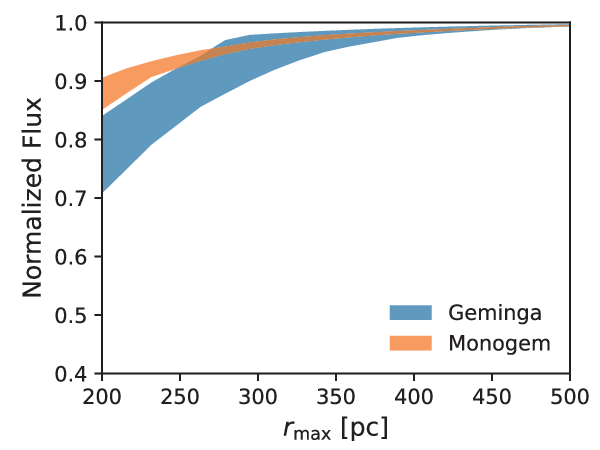}
    \caption{The flux within different integration radius to the total flux, considering the fitting result of $r_{\rm max}=200\,$pc, and a field of view radius of 10$^\circ$. The blue and orange bands are the 68\% confidence intervals for Geminga and Monogem, respectively.}
    \label{fig:normalized_flux}
\end{figure}

In Figure~\ref{fig:normalized_flux}, we calculated the ratio of flux within integration region to the total flux. The fractions are approximately 80\% and 90\% with $r_{\rm max}=200\,$pc for Geminga and Monogem, respectively. This indicate that an integration radius of 200\,pc is sufficient to contain most of the flux. And Table~\ref{tab:fit_max_dis} presents the fitting results for larger integration regions, showing no significant difference with that of $r_{\rm max}=200\,$pc.

\begin{table*}[h]
    \centering
    \setlength{\tabcolsep}{12pt}
    \begin{tabular}{cccccccccc}
    \hline
    \multirow{2}{*}{$r_{\rm max}$} & \multicolumn{4}{c}{Geminga} & & \multicolumn{4}{c}{Monogem}\\
    \cline{2-5} \cline{7-10}
    & $\phi\ [^\circ]$ & $\chi\ [^\circ]$ & $M_{\rm A}$ & $d\ [{\rm pc}]$ & & $\phi\ [^\circ]$ & $\chi\ [^\circ]$ & $M_{\rm A}$ & $d\ [{\rm pc}]$\\
    \hline
    300\,pc & $3^{+21}_{-3}$ & $7^{+28}_{-20}$ & $0.26^{+0.08}_{-0.13}$ & $389^{+160}_{-159}$ & & $15^{+6}_{-14}$ & $85^{+4}_{-16}$ & $0.15^{+0.04}_{-0.03}$ & $301^{+53}_{-47}$\\
    400\,pc & $1^{+5}_{-1}$ & $11^{+25}_{-23}$ & $0.22^{+0.11}_{-0.08}$ & $366^{+184}_{-169}$ & & $14^{+7}_{-13}$ & $85^{+4}_{-18}$ & $0.15^{+0.05}_{-0.03}$ & $302^{+53}_{-48}$\\
    500\,pc & $1^{+2}_{-1}$ & $12^{+24}_{-25}$ & $0.21^{+0.09}_{-0.07}$ & $351^{+191}_{-147}$ & & $14^{+7}_{-8}$ & $85^{+4}_{-10}$ & $0.15^{+0.04}_{-0.03}$ & $300^{+52}_{-47}$\\
    \hline
    \end{tabular}
    \caption{Medians and the 90\% confidence intervals obtained under the anisotropic diffusion model for Geminga and Monogem.}
    \label{tab:fit_max_dis}
\end{table*}

\bibliography{sample701}{}
\bibliographystyle{aasjournalv7}

\end{document}